\documentclass{aa}

\usepackage[utf8]{inputenc}
\usepackage{graphicx}
\usepackage{amssymb}
\usepackage{amsmath}
\usepackage{float}
\usepackage{natbib}
\bibpunct{(}{)}{;}{a}{}{,} 
\usepackage{array}
\usepackage{hyperref}
\usepackage{subfig}
\usepackage[varg]{txfonts}
\usepackage{multirow}
\usepackage{siunitx}
\usepackage{upgreek}
\usepackage{textgreek}

\usepackage{titlesec}

\usepackage{xcolor}

\hypersetup{colorlinks=true,citecolor=blue}

\titleformat{\paragraph}
{\normalfont\normalsize\bfseries}{\theparagraph}{1em}{}
\titlespacing*{\paragraph}
{0pt}{3.25ex plus 1ex minus .2ex}{1.5ex plus .2ex}

\def\app#1#2{%
  \mathrel{%
    \setbox0=\hbox{$#1\sim$}%
    \setbox2=\hbox{%
      \rlap{\hbox{$#1\propto$}}%
      \lower1.1\ht0\box0%
    }%
    \raise0.25\ht2\box2%
  }%
}

\DeclareTextSymbol{\degre}{T1}{6}
\DeclareTextSymbol{\degre}{OT1}{23}

\defcitealias{}{}          

\begin{document}
  \title{Search for water vapor in the high-resolution transmission spectrum of HD\,189733b in the visible}

   \subtitle{}

   \author{   R. Allart     \inst{1,*}, C. Lovis \inst{1}, L. Pino \inst{1,2}, A. Wyttenbach \inst{1}, D. Ehrenreich \inst{1}, F. Pepe \inst{1}
               }

   \institute{\inst{1} Observatoire astronomique de l'Universit\'e de Gen\`eve, Universit\'e de Gen\`eve, 51 chemin des Maillettes, CH-1290 Versoix, Switzerland\\
   					\inst{2} Dipartimento di Fisica e Astronomia `Galileo Galilei', Univ. di Padova, Vicolo dell'Osservatorio 3, Padova I-35122, Italy\\
              * \email{romain.allart@unige.ch}
             }

  \abstract
   {Ground-based telescopes equipped with state-of-the-art spectrographs are able to obtain high-resolution transmission and emission spectra of exoplanets that probe the structure and composition of their atmospheres. Various atomic and molecular species, such as Na, CO, H$_2$O have been already detected in a number of hot Jupiters. Molecular species have been observed only in the near-infrared while atomic species have been observed in the visible. In particular, the detection and abundance determination of water vapor bring important constraints to the planet formation process.}
   {We search for water vapor in the atmosphere of the exoplanet HD\,189733b using a high-resolution transmission spectrum in the visible obtained with HARPS. }
   {We use the atmospheric transmission code \texttt{Molecfit} to correct for telluric absorption features. Then we compute the high-resolution transmission spectrum of the planet using three transit datasets. We finally search for water vapor absorption in the water band around 6\,500\,\AA\ using a cross-correlation technique that combines the signal of 600 - 900 individual lines. }
   {Telluric features are corrected to the noise level. We place a 5-$\sigma$ upper limit of 100\,ppm on the strength of the 6\,500\,\AA\ water vapor band. The 1-$\sigma$ precision of 20\,ppm on the transmission spectrum demonstrates that space-like sensitivity can be achieved from the ground, even for a  molecule that is a strong telluric absorber.}
   {This approach opens new perspectives to detect various atomic and molecular species with future instruments such as ESPRESSO at the VLT. Extrapolating from our results, we show that only one transit with ESPRESSO would be sufficient to detect water vapor on HD\,189733b-like hot Jupiter with a cloud-free atmosphere. Upcoming near-IR spectrographs will be even more efficient and sensitive to a wider range of molecular species. Moreover, the detection of the same molecular species in different bands (e.g. visible and IR) is key to constrain the structure and composition of the atmosphere, such as the presence of Rayleigh scattering or aerosols (cloud and/or hazes).}

   \titlerunning{Search for water vapor in HD\,189733b with HARPS}
   \maketitle
%

\section{Introduction}
\begin{table*}[h]
\centering
\caption{\footnotesize Adopted physical and orbital parameters of HD\,189733b.}\label{paramètres utilisées}
\begin{tabular}{lccr}
\hline
Parameter & Symbol & Value & Reference \\
\hline
Stellar radius & $R_{*}$ & 0.756\,$\pm$\,0.018\,$R_{\odot}$ & \cite{torres_improved_2008} \\

Planet radius & $R_{p}$ & 1.138\,$\pm$\,0.027\,$R_{J}$ & \cite{torres_improved_2008} \\

White-light radius ratio & $R_{p}/R_{*}$ & 0.15617\,$\pm$0.00011\, & \cite{sing_hubble_2011} \\

Stellar mass & $M_{*}$ & 0.823\,$\pm$\,0.029\,$M_{\odot}$ & \cite{triaud_rossiter-mclaughlin_2009} \\

Planet mass & $M_{p}$ & 1.138\,$\pm$\,0.027\,$M_{J}$ & \cite{triaud_rossiter-mclaughlin_2009} \\

Epoch of transit & $T_{0}$ & 2454279.436714\,$\pm$\,0.000015\,BJD$_\mathrm{tdb}$ & \cite{agol_climate_2010}\\

Duration of transit & $T_{14}$ & 0.07527\,$\pm$\,0.00037\,d  & \cite{triaud_rossiter-mclaughlin_2009} \\

Orbital period & $P$ & 2.21857567\,$\pm$\,0.00000015\,d & \cite{torres_improved_2008}\\

Systemic velocity & $\gamma$ & -2.2765\,$\pm$\,0.0017\,km\,s$^{-1}$ & \cite{boisse_stellar_2009} \\

Semi-amplitude & K$_*$ & 200.56\,$\pm$\,0.88\,m\,s$^{-1}$ & \cite{boisse_stellar_2009}\\
\hline
\end{tabular}
\end{table*}

Over the past two decades, the field of exoplanets has expanded on a large scale with the development of numerous ground-based and space missions to detect and determine mass and radius of exoplanets with the radial velocity and transit techniques. Two of the most studied exoplanets, HD\,209458b \citep{charbonneau_detection_2000} and HD\,189733b \citep{bouchy_elodie_2005}, have been detected both by radial velocity and transit. These two exoplanets have a bright host star and are hot Jupiters, two characteristics that make them amenable to in-depth characterization. The aim of the characterization of an exoplanet is not only to determine its basic physical parameters, but also to determine its bulk composition and the composition of its atmosphere. To do so, transit techniques are available, either using the primary eclipse (for the transmission spectrum) or the secondary eclipse (for the thermal or reflected light). The first detection of atomic species was made on the transmission spectrum of HD\,209458b by \cite{charbonneau_detection_2002}, who detected the Na doublet in the visible with the STIS spectrograph on board the Hubble Space Telescope (HST). The first detection with a ground-based telescope was made by \cite{redfield_sodium_2008} on HD\,189733b with the Na doublet detection, then followed by \cite{snellen_ground-based_2008} for the same species in the atmosphere of HD\,209458b. These first ground-based sodium detections were made with slit spectrographs. Previous studies have emphasised the potential of stabilised and fiber-fed spectrographs for studying exoplanet atmospheres in the optical \citep{vidal-madjar_earth_2010,arnold_earth_2014}; this potential was demonstrated with HARPS for a hot gas giant by \cite{wyttenbach_spectrally_2015}. The presence of CO was detected in the infrared high-resolution transmission spectrum of HD\,209458b and of HD\,189733b by \cite{snellen_orbital_2010} and \cite{brogi_rotation_2016} using the CRIRES spectrograph at the Very Large Telescopes (VLT). \citet{deming_infrared_2013} and \citet{mccullough_water_2014} reveal the presence of H$_{2}$O in the infrared transmission spectrum of HD\,209458b and HD\,189733b with WFC3 instrument on HST. Today, tens of hot Jupiters have been characterized. Several studies indicate the presence of aerosols (clouds and/or hazes, e.g. \citealt{sing_continuum_2016}) through Rayleigh and Mie scattering signatures (e.g. \citealp{lecavelier_des_etangs_rayleigh_2008,sing_hst_2013,sing_hst_2015}), atmospheric evaporation (e.g. \citealp{lecavelier_des_etangs_evaporation_2010,bourrier_atmospheric_2013,ehrenreich_giant_2015}), temperature gradients within the atmosphere (e.g. \citealp{huitson_temperaturepressure_2012,wyttenbach_spectrally_2015,heng_non-isothermal_2015}), atmospheric circulation (e.g. \citealt{snellen_orbital_2010,kataria_atmospheric_2016,wyttenbach_spectrally_2015,louden_spatially_2015}) and an enhanced C/O ratio (e.g. \citealp{madhusudhan_high_2011,moses_chemical_2013,kreidberg_detection_2015}).\\
In this paper, we focus on HD\,189733b (see Table \ref{paramètres utilisées} for the physical and orbital parameters used). Its host star, HD\,189733, is an active, bright, metal-rich star of type K0{\sc V} (V-band magnitude of 7.65). The exoplanet is supposed to be a tidally-locked hot Jupiter and exhibiting a blue colour in the visible \citep{evans_deep_2013}. Its atmosphere contains Na \citep{redfield_sodium_2008,huitson_temperaturepressure_2012,wyttenbach_spectrally_2015,khalafinejad_exoplanetary_2016}, H$_{2}$O \citep{birkby_detection_2013,mccullough_water_2014,brogi_rotation_2016}, CO \citep{de_kok_detection_2013,brogi_rotation_2016} exhibits a Rayleigh scattering slope at blue wavelenghs \citep{pont_detection_2008,lecavelier_des_etangs_rayleigh_2008} likely caused by high-altitude hazes and loses hydrogen as shown by Ly-$\alpha$ absorption \citep{lecavelier_des_etangs_evaporation_2010,lecavelier_des_etangs_temporal_2012, bourrier_atmospheric_2013}.\\
In this study, we used data obtained with the HARPS high-resolution spectrograph ($\lambda/\Delta\lambda\sim$115\,000, \citealp{mayor_setting_2003}) to search for water vapor in the optical transmission spectrum of HD\,189733b. The high stability of HARPS allow us to optimally co-add hundreds of spectra while its high resolution allow us to resolve each individual line in the water spectrum which can be used to build a cross-correlation function (CCF) concentrating all the available water signal. So far, no detection of H$_{2}$O was ever made in the visible for an exoplanet, although weak water signatures are expected to be present. \\
In Sect.\,\ref{HARPS_obs} we describe the HARPS data used in this paper. In Sect.\,\ref{molecfit}, we correct these spectra from telluric features with the ESO tool Molecfit. In Sect.\,\ref{method}, we derive the transmission spectrum of the planet using the same method as \citet{wyttenbach_spectrally_2015,wyttenbach_hot_2017} and describe the cross-correlation function technique to study water vapor. Sect.\,\ref{result} shows the results which are then discussed in Sect.\,\ref{discussion}. We conclude in Sect.\,\ref{conclusion}.


\section{HARPS observations}

\label{HARPS_obs}
HD\,189733  was observed with HARPS mounted on the ESO 3.6m telescope at La Silla Observatory, Chile, in the programs 072.C-0488, 079.C-0127 (PI: Mayor) and 079.C-0828 (PI: Lecavelier des Etangs). These three programs contain four transits measured in 2006 and 2007 which are already analyzed by several authors.
These observations have yielded a number of important results: the study of the Rossiter-McLaughlin effect \citep{triaud_rossiter-mclaughlin_2009}, a 10-$\sigma$ detection of the Na doublet in the planet atmosphere \citep{wyttenbach_spectrally_2015}, winds circulating from the day side (hotter) to the night side (cooler) \citep{wyttenbach_spectrally_2015,louden_spatially_2015}, an atmospheric temperature gradient of 0.2-0.4\,K\,km$^{-1}$ \citep{wyttenbach_spectrally_2015,heng_non-isothermal_2015}, and a tentative of detection (2.5-$\sigma$) of Rayleigh scattering  \citep{di_gloria_using_2015}. Even the properties of the stellar surface occulted by the planet can be retrieved \citep{collier_cameron_line-profile_2010,cegla_rossiter-mclaughlin_2016}.\\
Table \ref{observation hd189733} provides the log of the observations used in this paper.
We note that an additional transit was observed on 29 July 2006 but only the first half of the transit was obtained due to bad meteorological conditions. Therefore, the transmission spectrum derived from this night is much more noisy than the other ones. As a consequence, we do not take it into account in the remainder of this paper. \\
The data reduction applied here was made with version 3.5 of the HARPS data reduction software (DRS). Spectra were extracted order by order (total of 72), flat-fielded using calibrations obtained at the beginning of the night, deblazed and wavelength calibrated. Finally, a one-dimensional spectrum from 3\,800 to 6\,900\,\AA\ with a step of 0.01\,\AA\ in the solar system barycentric rest frame was produced.

\begin{table}[h]
\centering
\caption{Observation log of the different nights showing the total number of spectra and the number of in- and out-transit spectra.}
\begin{tabular}{lcccc}
\hline
Date & 7 Sept 2006 & 19 Jul 2007 & 28 Aug 2007\\
\hline
Total spectra & 20 & 39 & 40 \\
In-transit & 11 & 19 & 19 \\
Out-of-transit & 9 & 20 & 21 \\
 t$_{exp}$ [s] & 600-900  & 300 & 300 \\
\hline
\end{tabular}
\label{observation hd189733}
\end{table}
		

\section{Correction of telluric contamination with \tt Molecfit}
\label{molecfit}
\begin{figure*}
\resizebox{\hsize}{!}{\includegraphics{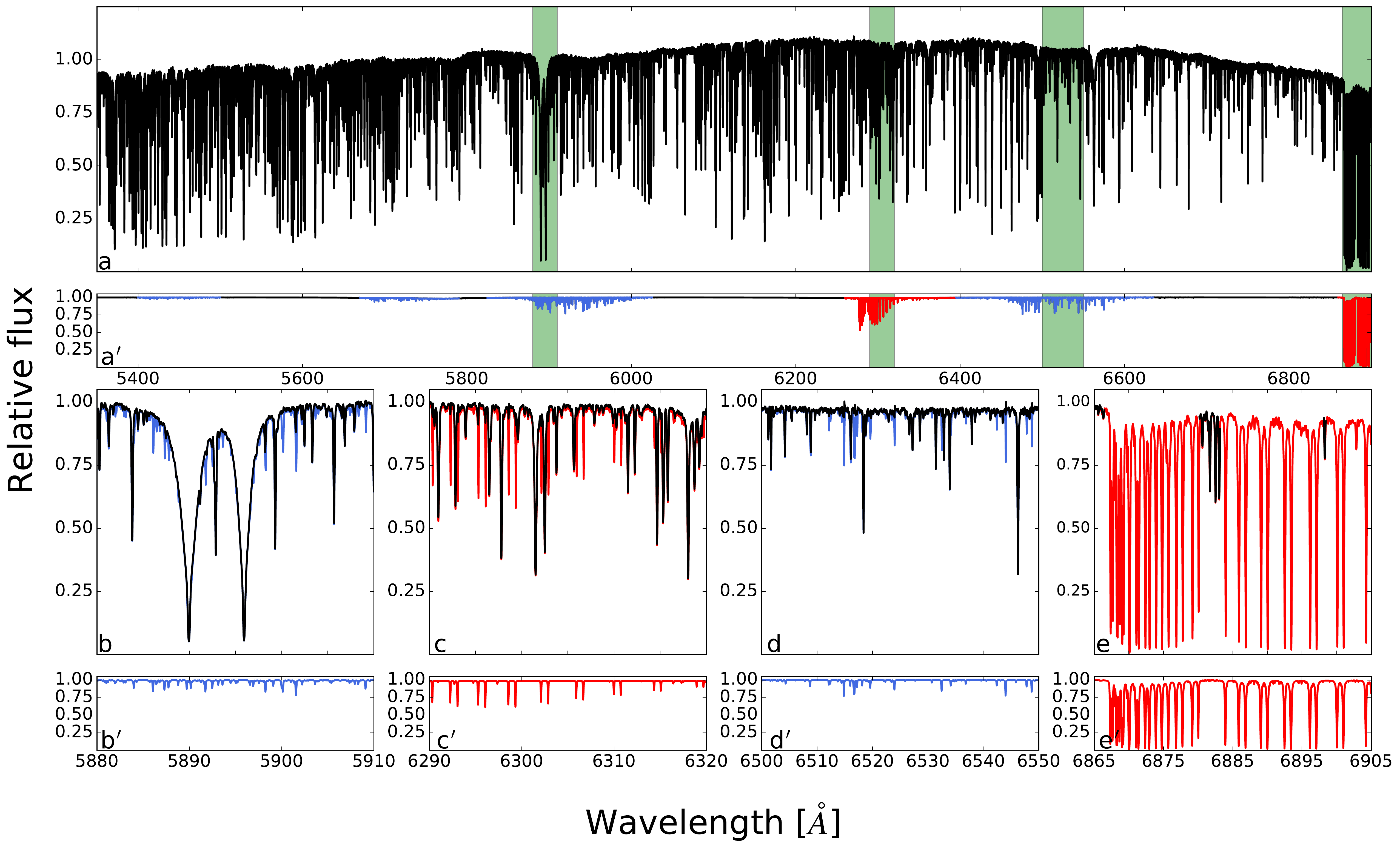}}
\caption[Influence des telluriques]{Influence of telluric lines on the spectrum of HD\,189733 obtained with HARPS on the red part of the CCD (5\,380 to 6\,900\,\AA) for the night of 19 July 2007. The upper panel (a, b, c, d, e) is the observed stellar spectrum with telluric contamination. Panel (a', b', c', d', e') is the best fit telluric model obtained with \texttt{Molecfit}  with H$_{2}$O in blue and O$_{2}$ in red. Panels (b, b'), (c, c'), (d, d') and (e, e') are close-up views of the green zones in panel (a,a'). These contain the strongest telluric bands in the visible (from left to right : H$_{2}$O (203-000, 302-000, 321-000) band at $\sim$5\,900\,\AA\ around the Na doublet, O$_{2}$ $\gamma$ band at $\sim$6\,200\,\AA\ , H$_{2}$O (311-000) band at $\sim$6\,500\,\AA\ and O$_{2}$ B band at $\sim$6\,900\,\AA\ ).}
\label{tellurique}
\end{figure*}

 Fig. \ref{tellurique} shows the red part of a HARPS spectrum and the influence of telluric lines in our optical ground-based observations. As we can see from the close-up views, the depth of some water telluric lines is $\sim$20\,$\%$. As we will see in Sect. \ref{non-detection}, the depth of the planetary water lines is expected to be 30 - 50\,ppm in the visible, which implies that the telluric correction is a  crucial step to probe the transmission spectrum of an exoplanet atmosphere.
\\
To perform this correction, we used version 1.2.0 of \texttt{Molecfit}  \citep{smette_molecfit_2015,kausch_molecfit_2015}, an ESO tool to correct telluric features in ground-based spectra. \texttt{Molecfit}  uses a line-by-line radiative transfer model (LBLRTM) to create a telluric spectrum at a very high resolution ($\mathrm{\lambda/\Delta\lambda} \sim$4\,000\,000). The LBLRTM needs an atmospheric profile that describes temperature, pressure, humidity and abundance of molecular species as a function of altitude for a given observatory site at one particular time and at a given airmass. To create this atmospheric profile, \texttt{Molecfit} merged an atmospheric standard profile and a Global Data Assimilation System (GDAS) profile.  The standard profile, which is provided by the Reference Forward Model \citep{remedios_investigation_2001}, described pressure, temperature, molecular abundances (up to tens molecular species) as a function of altitude for a specific latitude (e.g. equatorial, mid-latitude, polar-latitude for day or night). GDAS profiles provided by the National Oceanic and Atmospheric Administration (NOAA), are dedicated to wheater forecast. They contained meteorological data set (pressure, temperature, relative humidity as a function of altitude) and are updated every three hours for specific locations. The resulting merging atmospheric profile can be described in two possible grid, a fixed grid and a natural grid. The first one describes the variation of temperature, pressure, humidity and abundance of H$_{2}$O and O$_{2}$ from 0 to 120\,km with a fixed number of layers (50), while the second one is more precise with 100 to 150 layers. In this paper, we used the second grid which is more sensitive and yields a better telluric correction. Then the model spectrum is fitted to the observed spectrum by adjusting the continuum, the wavelength calibration and the instrumental resolution.

\begin{figure*}[t]
\resizebox{\hsize}{!}{\includegraphics{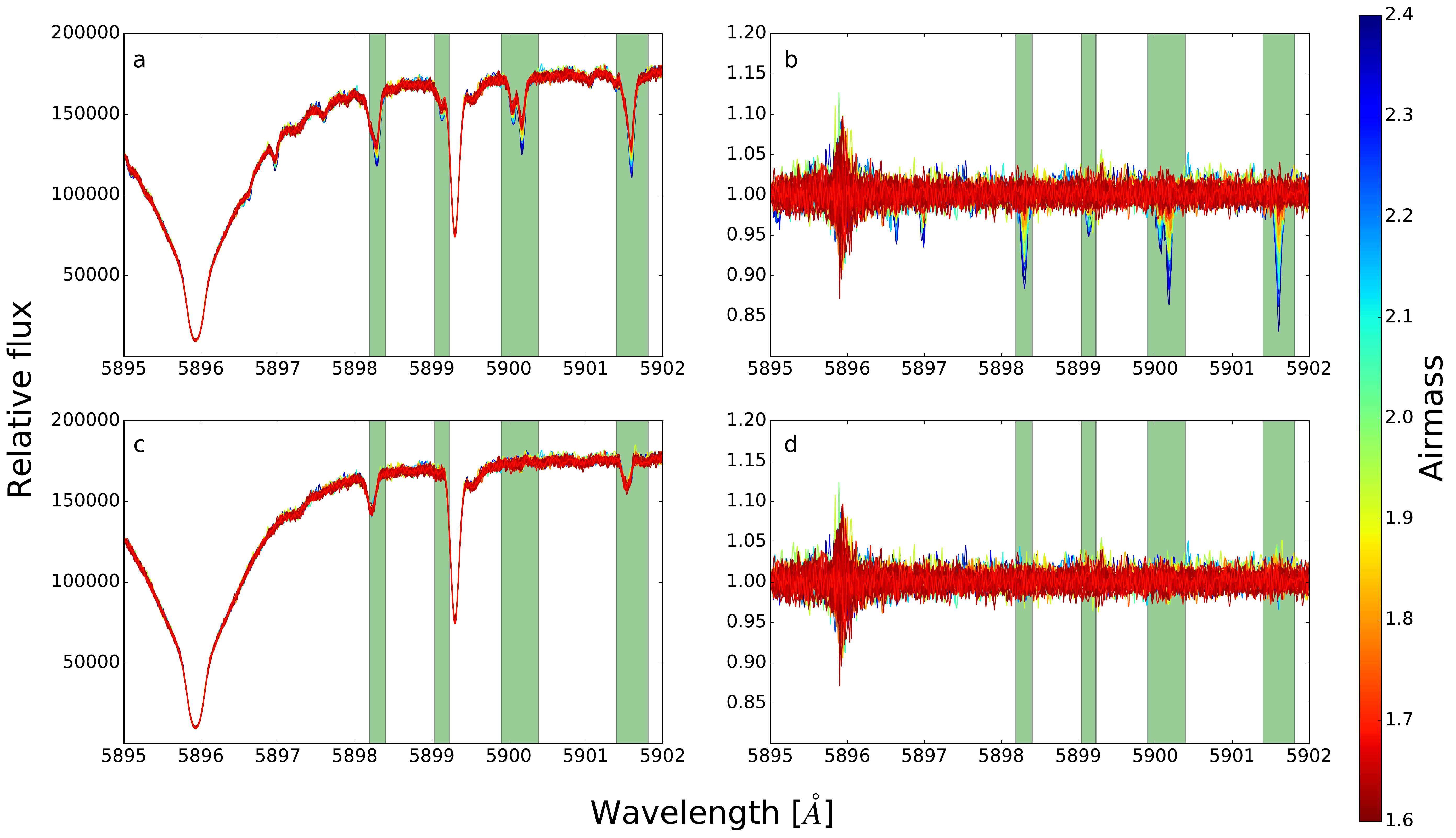}}
\caption[Résidu tellurique 5900 \AA]{Variability in the spectra over the night of 19 July 2007 for some water lines in the band at 5\,900\,\AA\ with the \ion{Na}{i} D1 line on the left. The colorbar indicates the airmass. Panel (a) shows the non-corrected spectra. Green bands indicate the strongest telluric lines. Panel (b) is the variation of each non-corrected spectrum from the mean spectrum of the night. Panel (c) shows spectra corrected from the telluric lines using \texttt{Molecfit}. Panel (d) is the variation of each corrected spectrum from the mean spectrum.}
\label{résidu tellurique 5900}
\end{figure*}

	\subsection{Adaptation of \textnormal{\texttt{Molecfit}}  to high-resolution visible spectra}
	
For the first time, \texttt{Molecfit}  is used on HARPS spectra, which are given in the Solar System barycentric rest frame, while the modelled spectrum by \texttt{Molecfit}  is given in the terrestrial rest frame. As a first step, we thus shifted the HARPS spectra into the terrestrial rest frame taking into account the Barycentric Earth Radial Velocity (BERV).\\
We then transformed the wavelength scale of the modelled spectrum from vacuum to air to match the observed spectrum. To optimise the correction of the telluric features, we decomposed the spectrum into about fifteen regions (part of them fall inside the green bands in Fig. \ref{tellurique}). We chose these regions such as to have only strong lines for a single molecule (H$_{2}$O or O$_{2}$), a flat continuum, and no stellar features within them. Indeed, \texttt{Molecfit}  does not model the stellar spectrum, and the fitting algorithm (Levenberg-Marquardt) is very sensitive to any stellar feature. We thus need to be very rigorous on the selection of spectral regions.\\
We refer to \cite{smette_molecfit_2015} for the detailed description of the free parameters of \texttt{Molecfit} . The goal of the fitting process is to adjust the continuum, the wavelength scale and the instrumental resolution for each fitted telluric band. The Levenberg-Marquardt $\chi^{2}$ convergence criterion and the parameter convergence criterion were both set to 10$^{-9}$.  The continuum was adjusted with a third-degree polynomial. The wavelength calibration was made with a Chebyschev second-degree polynomial. Finally, the instrumental profile  was assumed to be a Gaussian with a FWHM of 4.5 pixels. Appendix \ref{appendiceA} shows all the parameters used by \texttt{Molecfit} .\\
To correct the entire spectral range of interest, we used the Calctrans tool provided with \texttt{Molecfit} . It takes the best-fit parameters from the \texttt{Molecfit}  optimization and applies them to the entire spectrum. This operation was done for every individual spectrum in each night. The only difference in the settings used for each night consists in the choice of the fifteen spectral regions, which are optimized in each night with respect to the wavelength shift caused by the barycentric correction (Appendix \ref{appendiceB}).

	\subsection{A first assessment of the telluric correction}
\texttt{Molecfit}  produces an output file which contains all parameters of the fit. We explored a range of plausible parameter values to initialize the fitting process and verified that the fit converges to the same $\chi^{2}$ minimum, which it does. The obtained $\chi^{2}_{r}$ ranges from 3 to 12  for  68 to 80 parameters and 6\,760 to 10\,250 data points depending on the night. These relatively large values can likely be explained by the non-perfect model that is fitted to the data, in particular the non inclusion of stellar lines in the model.\\
Fig. \ref{résidu tellurique 5900} shows all the spectra for the night of 19 July 2007  before and after the telluric correction. The scatter in the Na doublet is due to the low flux in the core while the variability of water telluric features is due to changes in the water column density (mainly due to airmass variations). As we can see with panel (c), all the telluric lines are corrected to the noise level (including the telluric lines blended with stellar lines).


\section{Methods}
\label{method}
In this section, we describe how the transmission spectrum is derived and how we can search for water vapor in it with the cross-correlation technique.
	\subsection{Transmission spectrum}
The transmission spectrum \citep{seager_theoretical_2000,brown_transmission_2001} is obtained during the transit, when the planet passes in front of the star. It is defined as the area occulted by the planet over the stellar disk as a function of wavelength and probes the highest layers of the atmosphere.
We computed the transmission spectrum following the same formalism as  \cite{wyttenbach_hot_2017}.\\	
\begin{itemize}
\item[$\bullet$]Once the spectra are corrected from telluric features, we separated in-transit spectra from out-transit spectra. An in-transit spectrum, $f(\lambda,t_\mathrm{in})$, is a spectrum obtained when the planet occults a part of the stellar disk, while an out-transit spectrum, $f(\lambda,t_\mathrm{out})$, corresponds to the full-disk stellar spectrum.
\item[$\bullet$] We corrected each spectrum for the stellar reflex motion induced by the planet, using the orbital parameters in Table \ref{paramètres utilisées}.
\item[$\bullet$]Then, we created the normalized master out-transit spectrum which is the sum of the out-transit spectra:
\begin{eqnarray}
\tilde{F}_{\mathrm{out}}(\lambda)=\sum f(\lambda,t_\mathrm{out})
\label{master out}
\end{eqnarray}
\item[$\bullet$] We normalized all the in-transit spectra to the continuum level of the master out-transit spectrum with a fourth-degree polynomial. This normalization was made from 5\,338\,\AA\ to 6\,900\,\AA, which corresponds to the red CCD. We applied a sigma-clipping rejection algorithm on the spectra in order to replace all the cosmic ray hits by the mean value of the other spectra at each wavelength. A normalized in-transit spectrum is noted $\tilde{f}(\lambda,t_\mathrm{in})$.\\
\item[$\bullet$]Then, we computed the transmission spectrum as the sum of each individual transmission spectrum ($\tilde{f}(\lambda,t_\mathrm{in})/\tilde{F}_{\mathrm{out}}(\lambda)$), after applying a Doppler shift to the planet rest frame $p$ to compensate for the planet orbital motion during transit:
\begin{eqnarray}
\tilde{\mathfrak{R}}(\lambda)=\sum_{t\,\in\,\mathrm{in}} \left. \dfrac{\tilde{f}(\lambda,t_\mathrm{in})}{\tilde{F}_\mathrm{out}(\lambda)}\right|_{p}
\label{TS basic flux}
\end{eqnarray}

\item[$\bullet$]Finally, we expressed the variation of the occulted area by the planet as a function of wavelength on an absolute scale, by using the known white-light radius ratio (Table \ref{paramètres utilisées}) :
\begin{eqnarray}
\dfrac{R_{p}^{2}(\lambda)}{R_{*}^{2}}=1-\tilde{\mathfrak{R}}(\lambda)+\dfrac{R_{p}^{2}(\lambda_{\mathrm{ref}})}{R_{*}^{2}}
\label{TS basic surface}
\end{eqnarray}
\end{itemize}

	\subsection{Cross-correlation function}
	
The study of water vapor in the transmission spectrum in the visible is difficult considering the low intensity of the water bands and the noise level, even if we are able to resolve each line. To detect water vapor we need to co-add hundreds of lines to maximize the signal-to-noise ratio (S/N). To do so we used the cross-correlation function (CCF), expressed by :
\begin{eqnarray}
CCF(v)=\sum_{i} S(\lambda_{i}) \cdot M(\lambda_{i}(1+v/c))
\label{equation CCF}
\end{eqnarray}
This technique \citep{baranne_1996,pepe_2002_coralie} consists of projecting a binary mask $M$ (with an aperture width of one pixel) on the spectrum $S$ and to sum the transmitted flux at each wavelength. The mask contains the theoretical wavelengths of water vapor transitions, and is Doppler shifted successively to scan a radial velocity range of -80 to 80\,km\,s$^{-1}$ with a step of 0.82\,km\,s$^{-1}$. This step corresponds to the size of one pixel on the CCD of HARPS.\\
In this study, we used masks at five different temperatures for the water band between 6\,400 and 6\,800\,\AA , including one at 296\,K to check the presence of telluric residuals. The wavelength of each line was retrieved from HITRAN for the telluric mask (296\,K) and from HITEMP for exoplanetary masks (1\,300 to 2\,300\,K) \citep{rothman_hitran_2009,rothman_hitemp_2010}. The cross-section of each transition was scaled according to the temperature using the formulae provided in HITRAN. The number of lines in each mask was optimized to obtain the best S/N on the CCF. On the one hand, if the mask contains too few lines, even if they are strong, the noise will be large compared to the atmospheric signal. On the other hand, if too many lines are included, the noise will be lower but the weak lines will decrease the average signal. We found an optimal number of lines of 151-873 depending on temperature but note however that S/N has only a weak dependence on the exact line cut-off.
Table \ref{masque stellaire} listes the different water vapor masks with their spectral range and number of lines. The mask at 296\,K is used to check for the presence of telluric residuals. All the others are used to search for water vapor in the exoplanet.
\begin{table}[h]
\centering
\caption{Spectral range and number of lines used for each mask. 55\,226 water lines are available for this spectrale range in HITRAN and HITEMP database.}
\begin{tabular}{lcc}
\hline
Mask & Spectral range [\AA] & Nbr of lines used \\
\hline
$\mathrm{H_{2}O_{296\,K}}$&6\,424.92 - 6\,612.53&151 \\
$\mathrm{H_{2}O_{1300\,K}}$&6\,415.08 - 6\,763.10&239 \\
$\mathrm{H_{2}O_{1700\,K}}$&6\,400.63 - 6\,794.76&873 \\
$\mathrm{H_{2}O_{2100\,K}}$&6\,400.93 - 6\,794.76&606 \\
$\mathrm{H_{2}O_{2300\,K}}$&6\,409.47 - 6\,859.60&413 \\
\hline
\end{tabular}
\label{masque stellaire}
\end{table}


\section{Results}
\label{result}
In this section, we present the transmission spectrum, a second verification of telluric correction and the search for water vapor.

	\subsection{The transmission spectrum}
\begin{figure}[t]
\resizebox{\hsize}{!}{\includegraphics{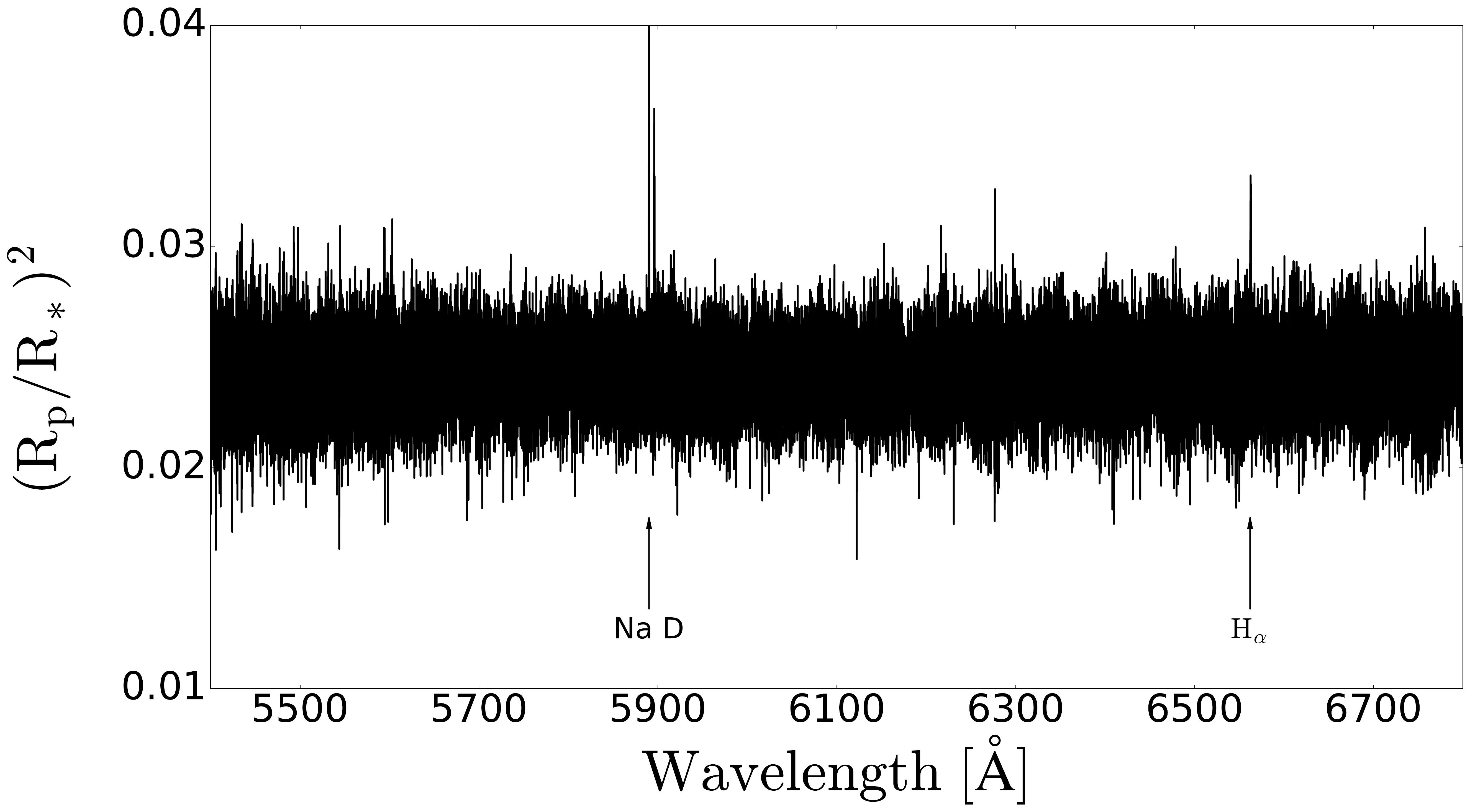}}
\caption[transmission spectrum]{The combined transmission spectrum of HD\,189733b. The Na doublet is visible at $\sim$5\,890\,\AA\ and the H-$\alpha$ line at 6\,562\,\AA. }
\label{transmission spectrum}
\end{figure}

In order to maximize the S/N, we built a weighted mean of our three transmission spectra (for the three nights) using $w_i=1/\sigma_{i}^{2}$ as weights, where $\sigma_{i}$ is the standard deviation in the continuum for each night $i$ . The transmission spectrum of HD\,189733b is shown in Fig. \ref{transmission spectrum} (standard deviation of 1\,500\,ppm). We confirm the excess absorption in the Na doublet at $\sim$5\,890\,\AA\ detected by \cite{wyttenbach_spectrally_2015}. Around 6\,562\,\AA , we can see the H$\alpha$ line, which varies with time (detected in two out of  three transits), and is a topic of much debate in the community \citep{barnes_origin_2016,cauley_optical_2015,cauley_variation_2016,cauley_evidence_2017}. We do not discuss it further here. 

	\subsection{The contribution of residual telluric features to the transmission spectrum}
	\label{CCF water 296k}
\begin{figure*}[h]
\resizebox{\hsize}{!}{\includegraphics{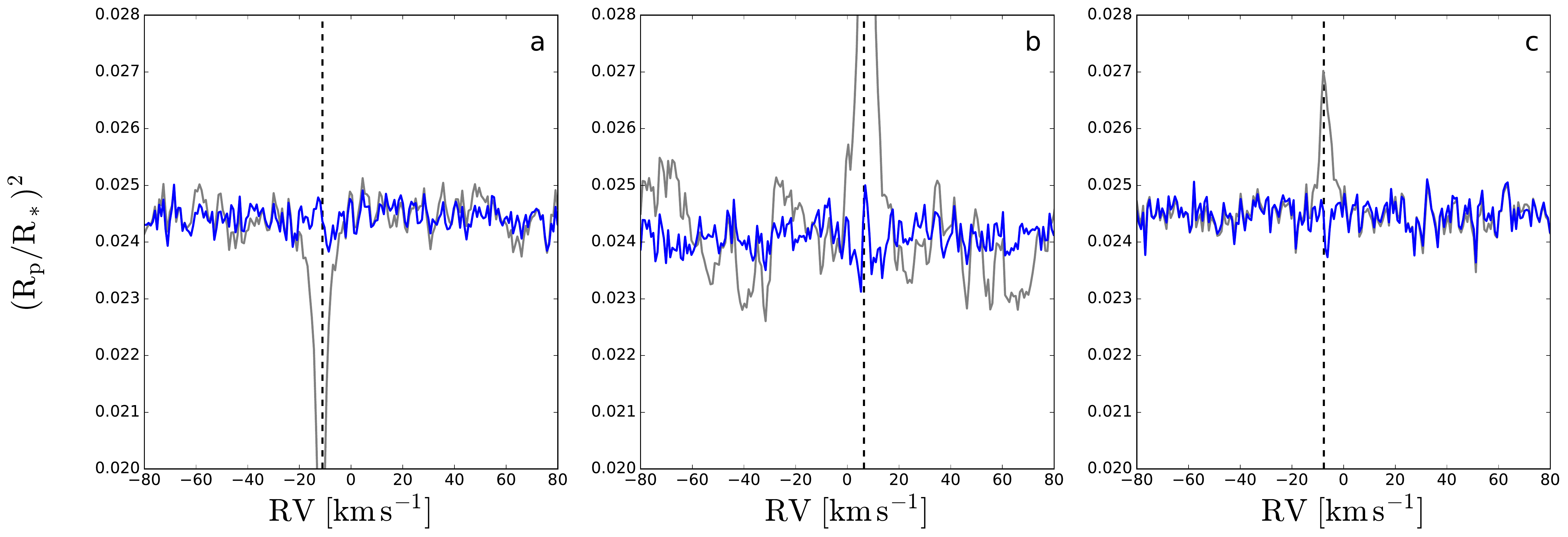}}
\caption[CCF eau 296K stellaire base]{CCFs of the transmission spectrum corrected from the telluric features (blue) and non-corrected (grey) in the stellar rest frame using the water vapor mask at 296\,K for the three nights (panel (a) : 7 September 2006, (b) : 19 July 2007  and (c) : 28 August 2007). The dashed line shows the observer's radial velocity, where telluric residuals would be expected.}
\label{CCF eau 296K stellaire base}
\end{figure*}
Fig. \ref{CCF eau 296K stellaire base} shows the CCF of the transmission spectrum corrected from telluric features (in blue) for the three nights obtained with the mask at 296\,K in the stellar rest frame. If the telluric correction was not sufficient, a signal would appear at the relative radial velocity of the terrestrial rest frame represented by the dashed line. The CCFs in grey are computed on the non-corrected transmission spectra and show the average telluric signature centered on the dashed line as expected. As we can see from the corrected CCFs, none of the nights shows a significant signal. The night of 19 July 2007 (panel b) shows somewhat higher dispersion in the core of the corrected CCF. Also, the non-corrected CCF shows a high dispersion in the continuum. This is due to  the different mean Doppler shifts of the telluric lines between the in- and out-transit spectra ($\sim \mathrm{300\,m\,s^{-1}}$). Indeed, the transit occurs at the begining of the night and thus the $\tilde{F}_{\mathrm{out}}(\lambda)$ is only composed with out-transit spectra taken after the transit.\\
As a conclusion, \texttt{Molecfit}  is capable of correcting the spectrum to the noise level  even when co-adding 151 telluric lines (dispersion in the CCF continuum of about $\sim$\,200\,ppm).
	\subsection{The search for water vapor in the transmission spectrum of HD\,189733b}
	\label{result water ccf}
\begin{figure}[h]
\resizebox{\hsize}{!}{\includegraphics{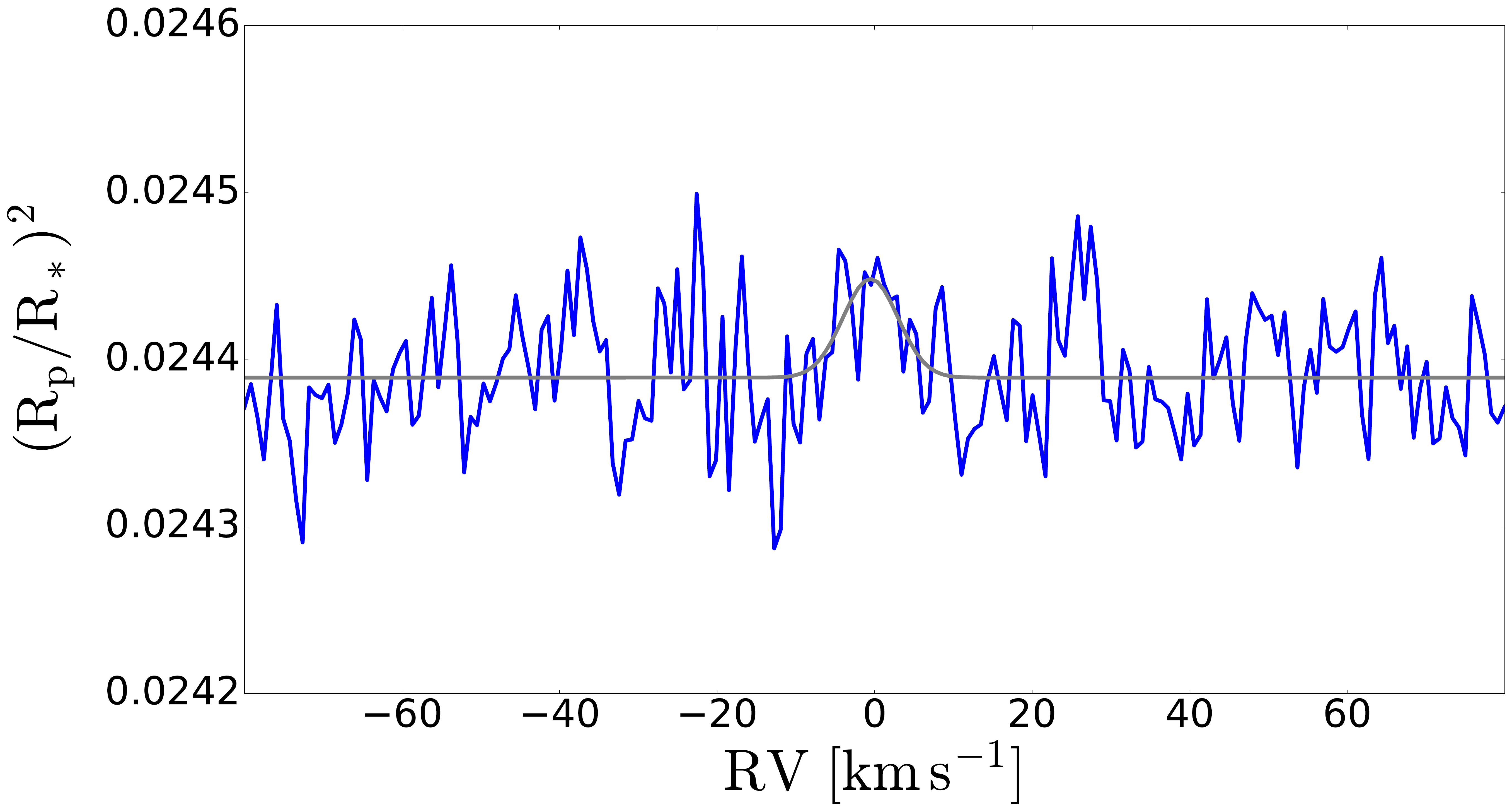}}
\caption[CCF eau 1700K planete base]{Measured CCF in the planet rest frame using the water vapor mask at 1\,700\,K (blue). The Gaussian fit is shown in grey.}
\label{CCF eau 1700K planet base}
\end{figure}

\begin{figure}[h]
\resizebox{\hsize}{!}{\includegraphics{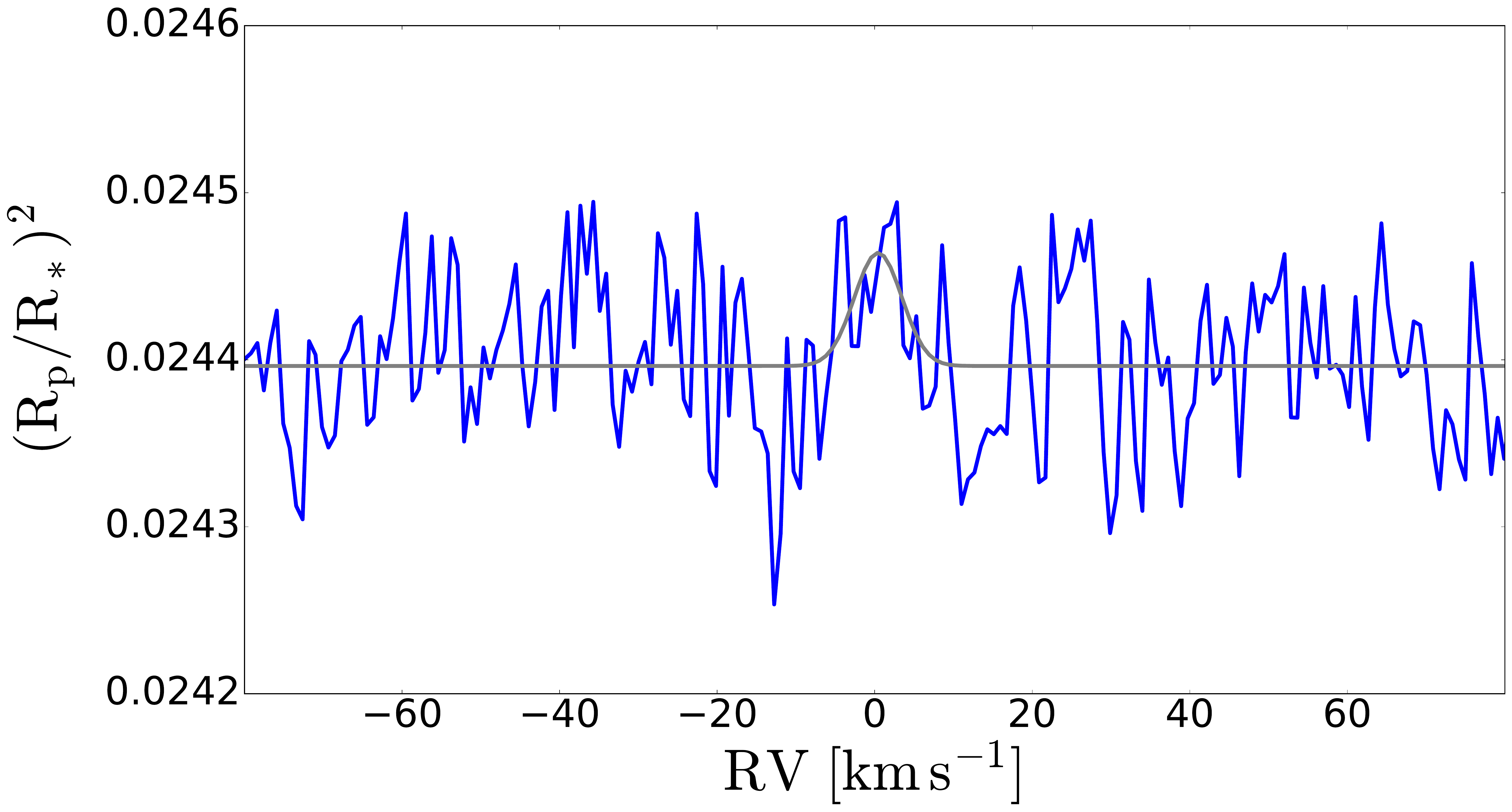}}
\caption[CCF eau 2100K planete base]{Measured CCF in the planet rest frame with the water vapor mask at 2\,100\,K (blue). The Gaussian fit is shown in grey.}
\label{CCF eau 2100K planet base}
\end{figure}

We built the water vapor CCF from the combined transmission spectrum using the masks at 1\,700\,K and 2\,100\,K. These temperatures yield the lowest noise (due to the number of lines) compared to the masks at 1\,300 and 2\,300\,K. We fit a Gaussian profile to the CCFs with free parameters for the continuum level, amplitude, FWHM and position. We also tested a Gaussian model with a FWHM fixed to 3.7 km\,s$^{-1}$ (Doppler broadening and instrumental profile) but found that the fitted parameters are within 1-$\sigma$ of the free FWHM parameters. Figs. \ref{CCF eau 1700K planet base} and \ref{CCF eau 2100K planet base} show the CCFs for the masks at 1\,700 and 2\,100\,K in the planet rest frame. The CCF at 1\,700\,K has a dispersion in the continuum $\sigma_{cont}$  of 34\,ppm. $\sigma_{cont}$ is defined as the standard deviation in the CCF continuum, ranging from -80 to -30\,km\,s$^{-1}$ and from 30 to 80\,km\,s$^{-1}$. The amplitude of the fitted Gaussian is 59 $\pm$ 17\,ppm, its centroid is at -0.54 $\pm$ 1.14\,km\,s$^{-1}$ and its FWHM is 8.2 $\pm$ 2.8\,km\,s$^{-1}$. The CCF at 2\,100\,K has a $\sigma_{cont}$ of 43\,ppm. The  amplitude of the fitted Gaussian is 68 $\pm$ 23\,ppm, its centroid is at 0.43 $\pm$ 116\,km\,s$^{-1}$ and its FWHM is 7.1 $\pm$ 2.8\,km\,s$^{-1}$. These $\sigma_{cont}$ values are due to white noise as they follow the square root of the number of lines (respectively 873 and 606).\\
\\
In spite of the formal results and uncertainties, we do not consider this as a significant detection of H$_2$O because similarly-good Gaussian fits can be obtained at other radial velocity positions (see Figs. \ref{CCF eau 1700K planet base} and \ref{CCF eau 2100K planet base}). Moreover, we carried out a model comparison between a Gaussian and a flat line using the Bayesian Information Criterion (BIC). We found that the Gaussian model is not optimal to fit these CCFs, since the straight line fit results in a $\Delta$BIC of 4 in favor of the straight line model. We thus think that the errors obtained by the square root of the covariance matrix are unreliable, because the $\chi^2$ minimum is poorly defined in parameter space.



\section{Discussion}
\label{discussion}
	\subsection{The telluric correction}
The telluric correction is one of the most critical steps to be able to study transmission spectra with ground-based facilities. In the visible, telluric lines are dominated by H$_{2}$O and O$_{2}$. We have shown that \texttt{Molecfit}  is a powerful tool which is able to correct H$_{2}$O telluric features to the noise level. However we stress here that one needs to be very careful on the choice of the fitted spectral regions.\\
One may ask if the telluric correction could have removed a part of the exoplanet water vapor signal. This is unlikely because telluric features can be easily distinguished from planetary water transitions thanks to their relative Doppler shift. Indeed, the relative radial velocity between the terrestrial and the planet rest frames varies from 0.4 to -31.3\,km\,s$^{-1}$ for the 7 September 2006, from 16.0 to -12.8\,km\,s$^{-1}$ for the 19 July 2007 and from 2.4 to -27.4\,km\,s$^{-1}$ for the 28 August 2007. Therefore, the two line systems overlap in only a small fraction of the observations. Moreover, \texttt{Molecfit} computes a physical model of the Earth atmosphere with essentially a single parameter controlling the depth of telluric water features. Also, the temperature on HD\,189733b is much higher than in Earth atmosphere, leading to different line intensities in the water band. Thus, even in the case of a null relative radial velocity, the telluric correction could erase only a fraction of the planetary water vapor signal. 
	\subsection{The non-detection of water vapor}
	\label{non-detection}
We computed a theoretical cloud-free isothermal transmission spectrum of HD\,189733b at the resolution of HARPS including H, He, Na, K and H$_{2}$O with solar abundances using the $^\pi \eta$ tool \citep{Pino_2017}. $^\pi \eta$ is an improved version of the $\eta$ code presented in \cite{ehrenreich_transmission_2006} and expanded in \cite{ehrenreich_transmission_2012} to compute transmission spectra of exoplanetary atmospheres. Its main characteristics are:
\begin{itemize}
\item[$\bullet$]   High-resolution (R $\sim$ 1\,000\,000). This is necessary to compare models to ground-based, high-resolution data.
\item[$\bullet$]  Broad wavelength coverage (330\,nm\,-\,2\,$\mu$m). This is necessary to compare models with space-borne, low- to medium-resolution data.
\end{itemize}

\begin{figure}[h]
\resizebox{\hsize}{!}{\includegraphics{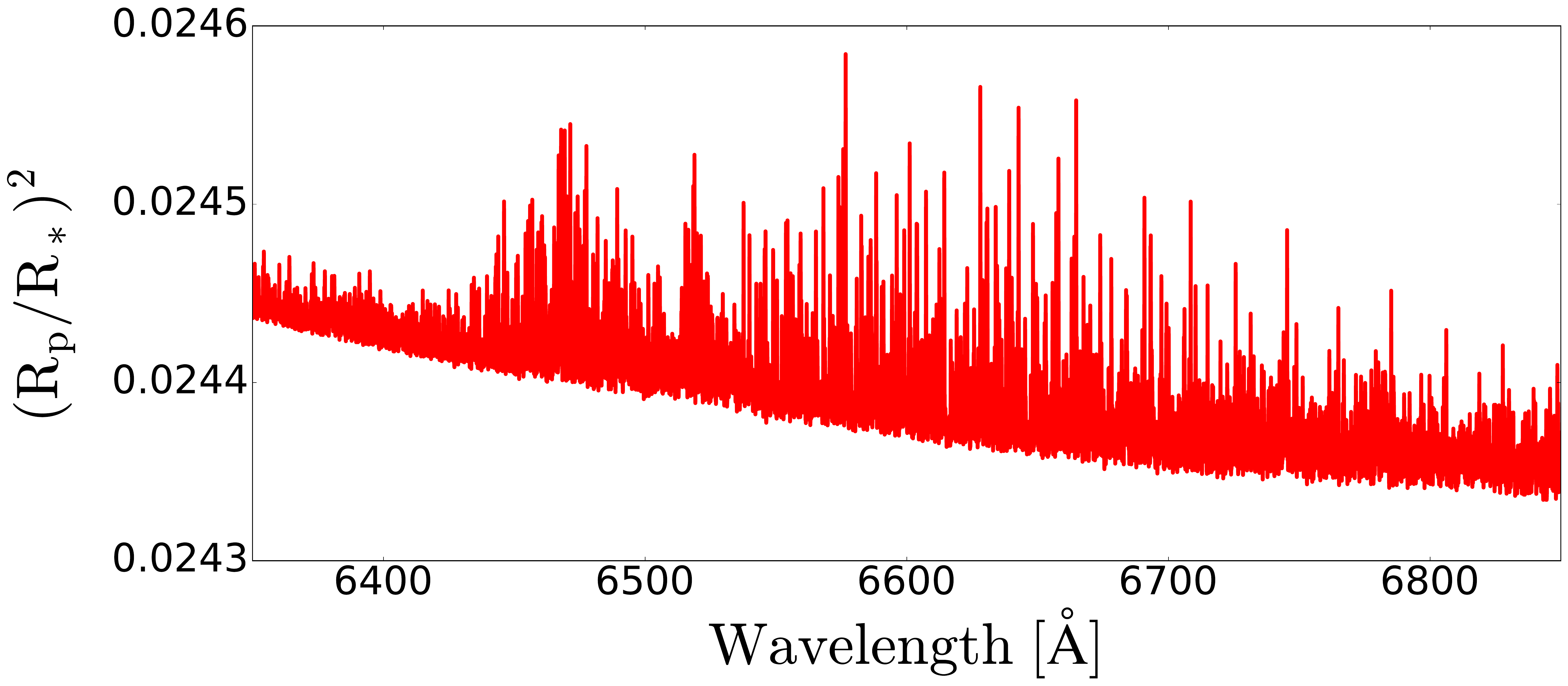}}
\caption[comparaison model and data]{Water band in the theoretical transmission spectrum at 2\,100\,K computed with $^\pi \eta$. The slope is caused by the red wing of the Na doublet.}
\label{TS theory 2100}
\end{figure}

\begin{figure}[h]
\resizebox{\hsize}{!}{\includegraphics{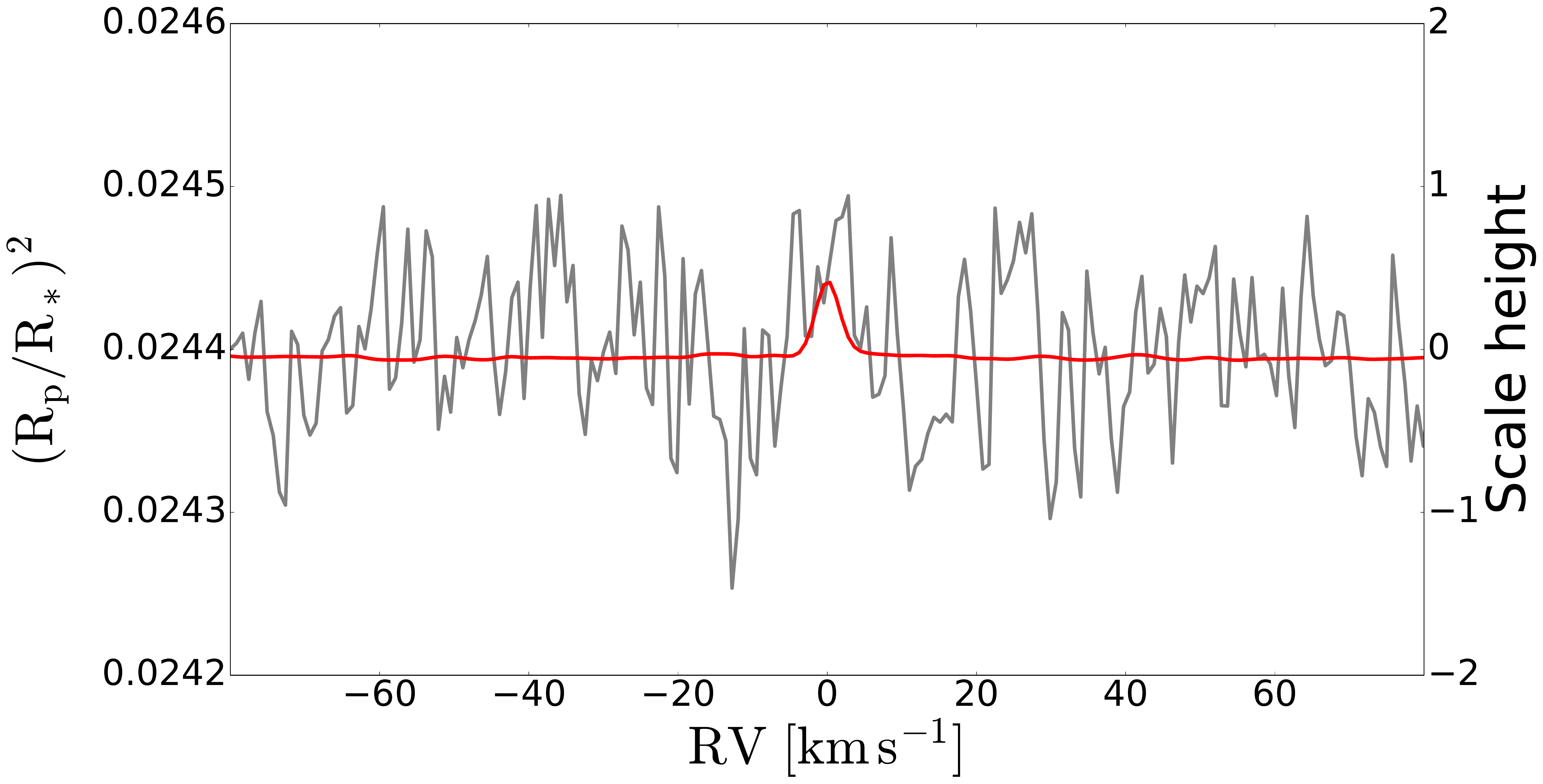}}
\caption[comparaison model and data]{Comparison between the measured CCF at 2\,100\,K in grey and the theoretical CCF at 2\,100\,K in red computed with $^\pi \eta$.}
\label{comparaison data and model}
\end{figure}

\begin{figure}[h]
\resizebox{\hsize}{!}{\includegraphics{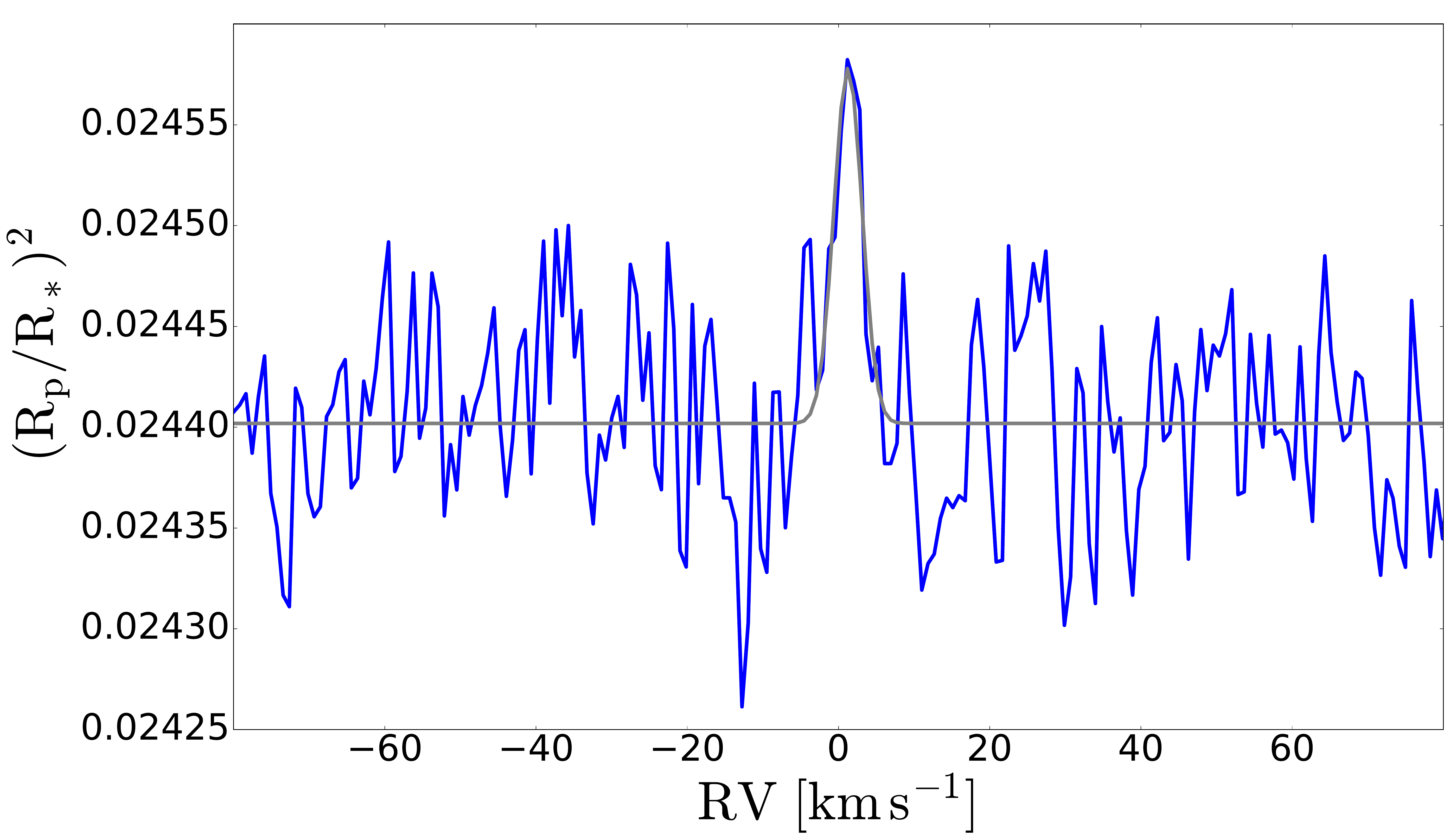}}
\caption[CCF eau 2100K planete base]{Measured CCF with an injected signal scaled to an average amplitude of 100\,ppm in the planet rest frame with the water vapor mask at 2\,100\,K (blue). The Gaussian fit is shown in grey.}
\label{CCF eau 2100K planet base injection}
\end{figure}

Fig. \ref{TS theory 2100} shows the 6\,500\,\AA\ water vapor band in the theoretical transmission spectrum at 2\,100\,K. We can thus compute the theoretical CCF with the two same water vapor masks that we used (1\,700 and 2\,100\,K). This is shown in Fig. \ref{comparaison data and model}. The predicted theoretical contrasts are 37 and 46\,ppm respectively, for a FWHM of 3.7 and 3.8\,km\,s$^{-1}$.\\
\\
As we can see in Fig. \ref{comparaison data and model}, the theoretical contrast is of the same order as the measured dispersion in the CCF continuum which is less than one half scale height. This implies that the present transmission spectrum has a noise level too high to detect water vapor in this planet. Table \ref{recap résultat} summarizes the noise properties and Gaussian fit parameters obtained from the data. To estimate our detection limits, we first compute the precision on the contrast $\sigma_{line}$ that could be achieved for a Gaussian CCF profile with a FWHM matching the theoretical value of 3.74\, km\,s$^{-1}$ (expressed in pixel units). It is given by :
\begin{eqnarray}
\sigma_{line}=\dfrac{\sigma_{cont}}{\sqrt{FWHM}},
\label{line noise}
\end{eqnarray}
where $\sigma_{cont}$ is the measured dispersion in the continuum. The 5-$\sigma$ detection limit is then simply given by 5$\cdot \sigma_{line}$. This gives a value of about 100\,ppm (see Table \ref{recap résultat}).\\
\begin{table}[h]
\centering
\begin{tabular}{lcc}
\hline
Mask temperature  [K] & 1\,700 & 2\,100 \\ 
Nbr of lines & 873 & 606 \\
\hline
$\sigma_{cont}$ [ppm] & 34 & 43  \\
$\sigma_{line}$ [ppm] & 16 & 20 \\
5-$\sigma$ detection limit [ppm] & 79 & 101\\
\hline
Fitted Gaussian contrast [ppm] & 59$\pm$17 & 68$\pm$23 \\
Theoretical contrast [ppm]& 37 & 46 \\
\hline
\end{tabular}
\caption{Main characteristics of the observed and theoretical CCFs at 1\,700 and 2\,100\,K.}
\label{recap résultat}
\end{table}
\\
To further test our detection limit estimates, we injected the model of Fig. \ref{TS theory 2100} scaled to an average amplitude of 100 ppm. The injection was done in the in-transit spectra before the telluric correction. We applied the same data reduction procedures as above. The Gaussian fit on the resulting CCF is shown in Fig. \ref{CCF eau 2100K planet base injection}. The amplitude of the fitted Gaussian is 176 $\pm$ 29\,ppm. The retrieved contrast is higher than the injected one but is still compatible with the noise level in the data (also given the uncertainty on the formal error itself, see discussion in Sect. \ref{result water ccf}). We also injected a 1\,000 ppm signal and retrieved a contrast of 1\,061 $\pm$ 29 ppm. This verification confirms that our technique does not remove the planetary signature.

	\subsection{Future prospects}
At this stage, the easiest way to detect water vapor in the atmosphere of HD\,189733b is by using the infrared transmission and emission spectra \citep{mccullough_water_2014,de_kok_detection_2013,brogi_rotation_2016}. The difficulty in the detection of the H$_2$O molecule in the visible comes from the weak water signature in this spectral band. Nevertheless, we can improve our S/N by using more observations and/or a more favourable water band. We can ask the question of how many transits with HARPS we would need to obtain a 5-$\sigma$ detection. Considering theoretical contrasts of 37 and 46\,ppm, a precision of 7 and 9\,ppm at 1\,700\,K and 2\,100\,K must be reached. This means that we would need to observe five times more transits than used in this study, so a total of 15 transits with HARPS.\\
ESPRESSO \citep{pepe_espresso_2010} is the new high-resolution spectrograph for the VLT, to be commissioned in 2017. ESPRESSO offers two major advantages compared to HARPS : first, the larger telescope diameter and improved instrument throughput will provide about six times more flux than HARPS on the 3.6\,m telescope. Second, the ESPRESSO spectral coverage extends to 7\,800\,\AA, which includes the potassium doublet at 7\,665 and 7\,699\,\AA, as well as a water band at 7\,400\,\AA\ . The latter is significantly stronger than the one studied in this paper. We computed with the $^\pi \eta$ code an expected CCF contrast of $\sim$ 200 ppm for a cloud-free atmosphere. Given these improvements, a detection could be obtained with just one transit. A succesful detection would allow us to compare the intensity of water absorption with the absorption features of the alkali doublets (sodium and potassium). By measuring their relative contrast, it is possible to infer the relative abundance of these species.\\
We need to keep in mind that the number of transits that we computed here is for a cloud-free atmosphere. If the atmosphere is cloudy as suggested by \citet{lecavelier_des_etangs_rayleigh_2008,pont_detection_2008,huitson_temperaturepressure_2012,pont_prevalence_2013,mccullough_water_2014}, water lines may be muted in these two bands. In the optical and NIR regions, scattering by aerosols is chromatic. Thus, the relative intensity of water bands at different wavelengths is indicative of the amount and type of scattering in the atmosphere. Spectral coverage is key to this characterization.
\label{harps to espresso}


\section{Conclusion}
\label{conclusion}
Our study can be summarized as follows:
\begin{itemize}
\item[$\bullet$] Telluric water lines are corrected to the noise level using the \texttt{Molecfit} tool.
\item[$\bullet$] We reach a 1-$\sigma$ precision of 20\,ppm on the CCF contrast for water vapor in the atmosphere of HD\,189733b. Therefore our data would have revealed a $\sim$100\,ppm signal at 5-$\sigma$.
\item[$\bullet$] Given a maximum theoretical contrast of only 46\,ppm, our data are too noisy to put a meaningful constraint on the presence of water vapor in HD\,189733b.
\item[$\bullet$] We determine how many transits we would need to detect water vapor in HD\,189733b-like planets with HARPS and the future ESPRESSO spectrograph for a cloud-free atmosphere. Given the increased efficiency and wavelength coverage of ESPRESSO, the stronger water band at 7\,400\,\AA\ could be detected in just one transit (CCF contrast of $\sim$\,200\,ppm). In conclusion, ESPRESSO will not only be a terrestrial planet hunter, but also the instrument of choice for atmospheric characterization in the visible.
\item[$\bullet$] The detection of a given species at different wavelengths (visible and near-IR) is key to characterize atmospheric structure and composition, in particular the presence of aerosols.
\end{itemize}


\begin{acknowledgements}
This work has been carried out within the frame of the National Centre for Competence in Research 'PlanetS' supported by the Swiss National Science Foundation (SNSF). The authors acknowledge the financial support of the SNSF by the grant numbers 200020\_152721 and 200020\_166227. We want to thank the entire atmosphere group of Geneva Observatory, the team in charge of \texttt{Molecfit}  for their work and their upgrades, Jens Hoeijmaker about the discussion on the water correction and the anonymous referee for
the careful reading and pertinent comments.
\end{acknowledgements}
\bibliographystyle{aa}
\bibliography{bib}

\begin{appendix}
\section{Appendix A}
\label{appendiceA}

\begin{table}[h]
\centering
\caption{Initial parameters of \texttt{Molecfit}  for every night.}\label{parametre molecfit}
\begin{tabular}{lll}
\hline
Initial parameters & Values & Commentary \\
\hline
ftol&$10^{-9}$& $\chi^{2}$ Convergence criterion \\
xtol&$10^{-9}$& Parameter convergence criterion\\
molecules&H$_{2}$O, O$_{2}$&  \\
$n_{cont}$ &3& Degree of polynom for the continuum \\
$a_{0}$&2000& Constant of polynom for the continuum\\
$n_{\lambda}$&2& Chebyschev degree for wavelength calibration\\
$b_{0}$&0& Constant Chebyschev for wavelength calibration\\
$\omega_{gaussian}$&4.5&FWHM in pixel\\
kernel size&15& \\
pixel scale&0.16& \\
slit width&1"& \\
MIPAS profil &equ& Equatorial profil\\
atmospheric profil&0& Natural profil\\
PWV &-1& No value taking into account\\
\hline
\end{tabular}
\end{table}

\section{Appendix B}
\label{appendiceB}

\begin{table}[h]
\centering
\caption{Fitted regions with \texttt{Molecfit} for the three nights.}
\begin{tabular}{ccc}
\hline
\multicolumn{3}{c}{Fitted regions [$\mathrm{\mu m}$]} \\
\hline
7 Sep 2006 & 19 Jul 2007 & 28 Aug 2007 \\
\hline
0.592100 - 0.592354 & 0.592040 - 0.592345 & 0.592100 - 0.592253 \\
0.592500 - 0.592927 & 0.592393 - 0.592906 & 0.592500 - 0.592703 \\
0.594676 - 0.594912 & 0.594530 - 0.594890 & 0.594716 - 0.594798 \\
0.596814 - 0.597700 & 0.596781 - 0.597663 & 0.596916 - 0.597095 \\
0.627961 - 0.628145 & 0.627882 - 0.628203 & 0.597258 - 0.597700 \\
0.628313 - 0.628410 & 0.628270 - 0.628405 & 0.628034 - 0.628111 \\
0.628510 - 0.628671 & 0.628470 - 0.628655 & 0.628307 - 0.628390 \\
0.628861 - 0.629231 & 0.628800 - 0.629232 & 0.628510 - 0.628671 \\
0.629614 - 0.629818 & 0.629578 - 0.629793 & 0.629064 - 0.629231 \\
0.647405 - 0.647726 & 0.647370 - 0.647711 & 0.629614 - 0.629818 \\
0.647964 - 0.648327 & 0.647790 - 0.648328 & 0.647465 - 0.647726 \\
0.648530 - 0.649333 & 0.648465 - 0.649300 & 0.648078 - 0.648233 \\
0.651200 - 0.652005 & 0.651100 - 0.651964 & 0.649058 - 0.649333 \\
0.686800 - 0.688240 & 0.686850 - 0.688216 & 0.651626 - 0.651717 \\
0.688550 - 0.691500 & 0.688509 - 0.691500 & 0.651869 - 0.652005 \\
                                  &                                   & 0.686684 - 0.687365 \\
                                  &                                   & 0.687521 - 0.688240 \\
                                  &                                   & 0.688539 - 0.691236 \\                                                       
\end{tabular}

\label{recap regions}
\end{table}

\end{appendix}
\end{document}